\documentclass{article}

\usepackage{PRIMEarxiv}
\usepackage{float}
\usepackage{amsmath}
\usepackage[utf8]{inputenc} 
\usepackage[T1]{fontenc}   
\usepackage{hyperref}       
\usepackage{url}          
\usepackage{booktabs}       
\usepackage{amsfonts}      
\usepackage{nicefrac}       
\usepackage{microtype}     
\usepackage{lipsum}
\usepackage{fancyhdr}      
\usepackage{graphicx}      
\graphicspath{{media/}}     

\title{ChemReservoir - An Open-Source Framework for Chemically-Inspired Reservoir Computing

\thanks{The work was supported by the European Union and the Swiss State Secretariat for Education, Research and Innovation (SERI) under contract numbers 22.00017 and 22.00034 (Horizon Europe Research and Innovation Project CORENET)} 
}

\author{
  Mehmet Aziz Yirik \\
  Department of Mathematics and Computer Science \\ University of Southern Denmark\\ 
  Campusvej 55, DK-5230 Odense, Denmark \\
  \texttt{mehmetazizyirik@gmail.com} \\
   \And
  Jakob Lykke Andersen, Rolf Fagerberg \\
  Department of Mathematics and Computer Science \\ University of Southern Denmark\\ 
  Campusvej 55, DK-5230 Odense, Denmark \\
  \texttt{\{jlandersen, rolf\}@imada.sdu.dk} \\
   \And
  Daniel Merkle \\
  Algorithmic Cheminformatics \\
  Faculty of Technology, Bielefeld University \\ Universitätsstraße 25, 33615 Bielefeld, Germany \\
  \texttt{daniel.merkle@uni-bielefeld.de} \\
}

\begin{document}

\begin{center}
\textit{This work has been submitted to the IEEE for possible publication. Copyright may be transferred without notice, after which this version may no longer be accessible.}
\end{center}

\maketitle

\begin{abstract}
Reservoir computing is a type of a recurrent neural network, mapping the inputs into higher dimensional space using fixed and nonlinear dynamical systems, called reservoirs. In the literature, there are various types of reservoirs ranging from in-silico to in-vitro. In cheminformatics, previous studies contributed to the field by developing simulation-based chemically inspired in-silico reservoir models. Yahiro used a DNA-based chemical reaction network as its reservoir and Nguyen developed a DNA chemistry-inspired tool based on Gillespie algorithm. However, these software tools were designed mainly with the focus on DNA chemistry and their maintenance status has limited their current usability. Due to these limitations, there was a need for a proper open-source tool. This study introduces ChemReservoir, an open-source framework for chemically-inspired reservoir computing. In contrast to the former studies focused on DNA-chemistry, ChemReservoir is a general framework for the construction and analysis of chemically-inspired reservoirs, which also addresses the limitations in these previous studies by ensuring enhanced testing, evaluation, and reproducibility. The tool was evaluated using various cycle-based reservoir topologies and demonstrated stable performance across a range of configurations in memory capacity tasks. 
\end{abstract}

\keywords{Chem-Inspired Computing \and Machine Learning \and Reservoir Computing}

\section{Introduction}
\label{sec:introduction}
Machine learning (ML) is a broad term comprising various methods such as neural networks and kernel methods. Among ML methods, recurrent neural networks (RNNs) have been widely used for tasks such as classification and prediction of time series. Despite the efficient performance of RNNs, the computational cost of backpropagation and the long training times have triggered the development of more energy efficient RNN models \cite{shahi2022prediction}. One of the earliest attempts was by Buonomano \cite{buonomano1995temporal}, who implemented a random network of spiking neurons in which only the output layer was trained. Following Buonomano’s study, Jaeger refined the idea and contributed to the field of Echo State Networks (ESNs)\cite{jaeger2002tutorial}.  This RNN type is called the reservoir computing model. It comprises three layers -input, reservoir, and readout layers- to map input data into a high-dimensional, nonlinear feature space called the reservoir to capture complex input patterns. Typically, linear or regularized linear (ridge) regression is performed in the readout layers. The primary distinction between traditional recurrent neural networks (RNNs) and reservoir computing networks is the training process. In RNNs, edge weights across network components must be trained. In contrast, reservoir computing models require only training in the readout layer. The key feature of an efficient reservoir is its underlying topology, which should exhibit rich dynamics to map to a high-dimensional space. The rich dynamics also has an impact on the memory capacity of the reservoir to store input history and spatio-temporal patterns \cite{zhang2023survey}. Many reservoir models possess short-term memory, where the current output depends on previous inputs \cite{jaeger2002tutorial}. The length of this memory is determined by the dynamics and topology of the reservoir. The entry of new inputs into the reservoir shifts the dynamics, causing fading memory. The fading memory is a key characteristic of an efficient reservoir model, as this allows the reservoir to capture input-specific features. The essential criteria that define an efficient reservoir are captured by the Echo State Property (ESP) \cite{lukovsevivcius2009reservoir}: 

\begin{itemize}
  \item Encoding of spatial-temporal information from input data in immediate reservoir states.
  \item Retention of relevant information from past inputs over time, with a gradual decline in memory retention (fading memory).
\end{itemize}

The design of ML architectures consists not only of algorithmic strategies but also of the integration of physical systems. Recent studies in ML, especially neuromorphic computing, exhibit the potential impact of physical systems, such as chemical reaction networks, in reservoir models \cite{csizi2024complex, baltussen2024chemical}. For example, Baltussen et al. \cite{baltussen2024chemical} modeled an in-vitro formose chemistry based-chemical reservoir model which can efficiently perform classification tasks. In addition to in-vitro studies, there have been attempts for the design of chemically-inspired reservoir (CIR) models, where simulation-based in-silico reservoirs were designed based on DNA chemistry 
\cite{nguyen2020reservoir, yahiro2018reservoir}. However, these two software are 
not actively maintained and benchmarking was not conducted rigorously, raising concerns 
about the reliability of the results. Additionally, Yahiro et al. 
\cite{yahiro2018reservoir} mentioned that over-fitting could be a problem in their 
results and Nguyen et al.'s \cite{nguyen2020reservoir} training process could be 
improved by a better division between training and test data sets. These preliminary in-silico trials -and their limitations- underscore the need for robust frameworks to investigate chemically inspired reservoir computing systems. In this study, ChemReservoir, an open-source framework for CIR computing is introduced for the in-silico construction and conceptual analysis of CIR models. Previous studies have evaluated cycle-based topologies and reported their effectiveness in preserving the rich reservoir dynamics \cite{yahiro2018reservoir, rodan2010minimum}. Therefore,  ChemReservoir was also tested with cycle-based topologies, where various cycle lengths, numbers of chords, and input network parameters were evaluated using genetic algorithms. ChemReservoir comprises two genetic algorithms to optimize the topology selection and network parameter selection for the optimal network features. The chords are added to the topologies to introduce a feedback mechanism preserving an 'echo' of recent input while limiting the influence of past inputs, thereby supporting the echo state property. ChemReservoir topologies consist of pseudo-
molecules as nodes and pseudo-rules as edges (abstract analogs of molecules and 
reaction rules) within the network. Memory capacity benchmarks are widely utilized 
to assess reservoir models. Consistent with previous studies \cite{yahiro2018reservoir, nguyen2020reservoir}, we also evaluated memory capacity using short- and long-term memory tasks, which are simplified versions of the widely used NARMA benchmark in reservoir computing \cite{wringe2025reservoir}. The input signals were generated based on random numbers sampled from a normal distribution and input values are fixed in the signal for specific input hold time referred to as steps. Our results show that reducing the step size or increasing the frequency of changes in the inflow values led to better reservoir 
performance in memory capacity tasks. This result is attributed to the reservoir's 
dynamic response to more frequent changes in the input signal, which also has an 
impact on the fading memory of the reservoir models. In the short-term memory task, 
we observed that the density of chords influenced the test errors. Our results from the long-term memory task demonstrate that the increase in the past time lag led 
to higher learning errors, which is expected due to the fading memory 
characteristic of the reservoir model. Overall, ChemReservoir shows promising 
results with respect to the echo-state property and provides an open-source 
framework for chemically inspired reservoir models.

\section{Methods}
\label{sec:methods}

\subsection{Construction and Simulation of Reservoir Topologies}
\begin{figure}[H]
  \centering
  \includegraphics[width=0.4\linewidth]{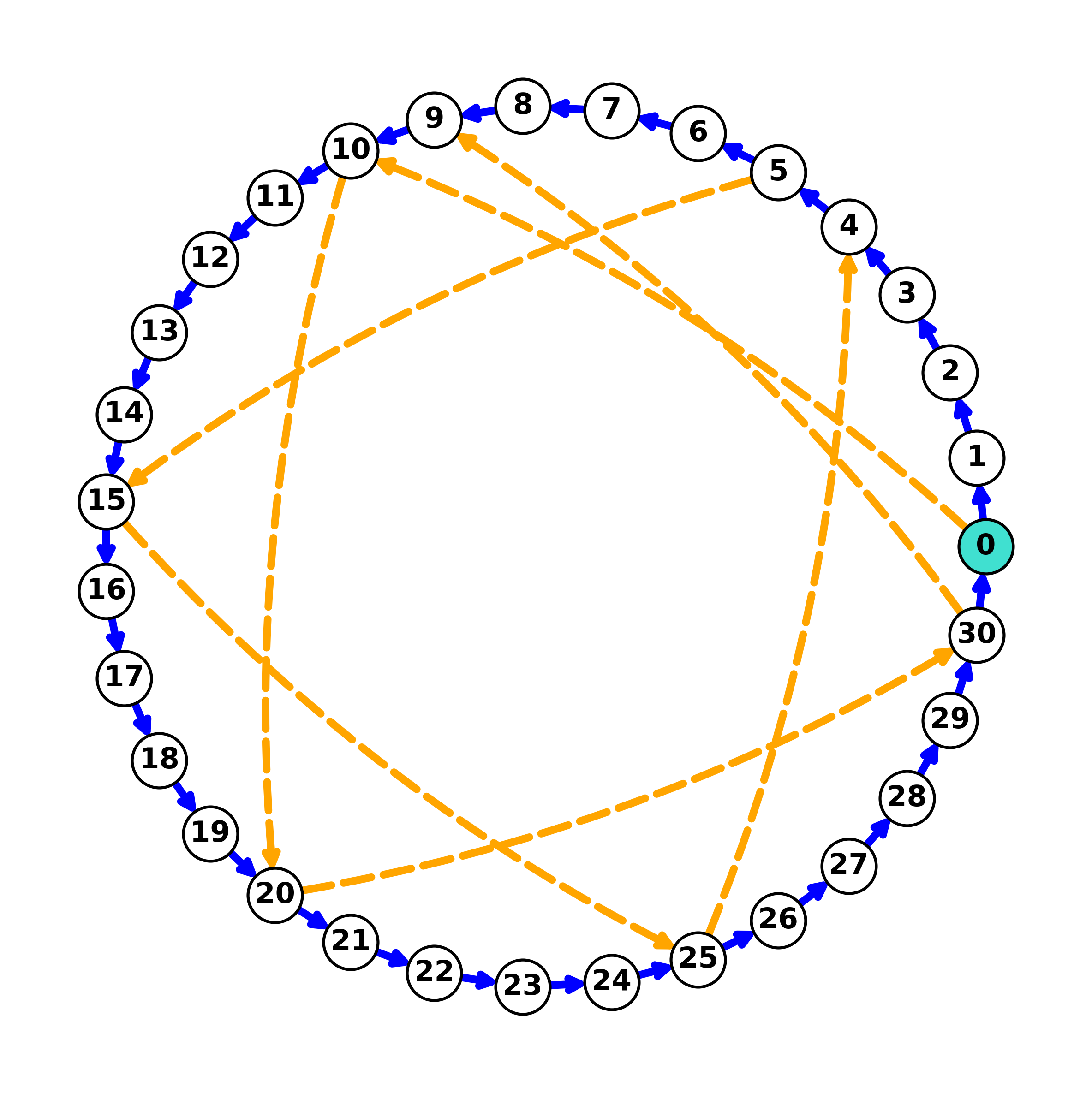}
  \caption{Reservoir network with 30 nodes, chord length 10, and chord step size 5. Cycle edges are colored blue, chords in orange. Inflow is fed via node 0, colored light blue.}
  \label{fig:exampleNetwork}
\end{figure}

MØD \cite{andersen2016software} software is utilized for the construction of abstract reaction networks -conceptually inspired by chemical reaction networks- as the underlying topology of CIRs. Their design involves defining the pseudo-molecules and pseudo-reaction (abstract analogs of molecules and reaction rules), as the nodes and edges of the reservoir topology. Cycle-based reservoir systems, which have been shown to enhance reservoir dynamics \cite{yahiro2018reservoir, rodan2010minimum}, are utilized as the foundational topology of ChemReservoir. In these cycle-based topologies, a variety of chords are added to assess and enhance local connectivity within the reservoirs. Local connectivity is crucial for long-range temporal dependencies to preserve information from earlier inputs. Therefore, chords act as feedback loops, facilitating information propagation and increasing nonlinearity in reservoir dynamics. An example reservoir topology is illustrated in Fig. \ref{fig:exampleNetwork}.

These networks are used as the underlying reservoir topologies. In ChemReservoir, an input signal is fed into the network via a food molecule (indexed as 0) and the reservoir undergoes a stochastic simulation for a specific duration. The stochastic simulation results -the molecule counts for each node- are stored to be used later for the training phase. Therefore, in addition to the structural definition, parameters for the stochastic simulation -such as the amount of inflow and the rule rates- must also be defined for each topology. A recently developed MØD module, called MØD-StochSim, is utilized for the stochastic simulation of reaction networks. MØD-StochSim is an implementation of the Gillespie algorithm, a widely implemented approach for stochastic simulation of chemical reaction systems \cite{effenberger2022biology}. The tool serves as the stochastic simulation component within ChemReservoir. The method models the occurrence of different events in a chemical reaction over time, namely input, output, and reaction events. For stochasticity, the events are randomly selected during the simulation on the basis of quantities of molecules present in the system. The parameters used to construct the topologies and simulations are explained in detail in Section \ref{subsec:params}. 

\subsection{Construction of CIR Model}

\begin{figure}[H]
  \centering
  \includegraphics[width=0.8\linewidth]{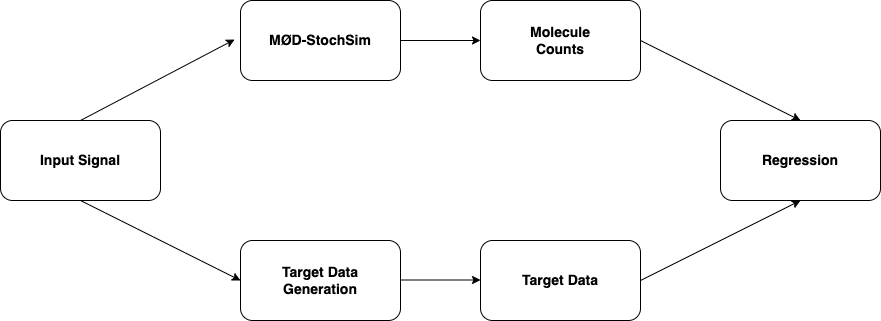}
  \caption{The ChemReservoir Workflow.}
  \label{fig:modChart}
\end{figure}

A CIR model consists of three layers: input, stochastic simulation, and readout layer. In this model, the underlying reservoir network is simulated for a period of time for input parameters and inflow. The simulation data are stored as the molecule count over time for each pseudo-molecule. In parallel, the same inflow is used for the generation of target data. The stored stochastic simulation data and the target data are input for the regression step. In CIR models, the ridge regression method \cite{pedregosa2011scikit} is performed in the readout layer. Fig. \ref{fig:modChart} depicts the workflow of a CIR model.

\subsection{Memory Capacity Tasks}

Memory capacity tasks are widely used benchmarks in reservoir computing, especially the simplified versions of the NARMA benchmark, such as short-memory and long-memory tasks  \cite{wringe2025reservoir}. These tasks evaluate the memory capacity of reservoir models based on their learning performance in the regression step. This also provides a way to assess the fading memory of the reservoir during the learning process \cite{lukovsevivcius2009reservoir}. Yahiro and Nguyen also performed short- and long-term memory tasks to benchmark the memory capacity of the reservoir for the given input signals. ChemReservoir was also evaluated using memory capacity benchmarks. The short-term memory target data are defined on the basis of the following formula \eqref{eq1}:

\begin{equation} Y'(t)= Q^k_{in}(t-1) + 2*Q^k_{in}(t-2)\label{eq1}\end{equation}

$Y'(t)$ is the target data at time t. $Q^k_{in}(t)$ is the influx value of the $k^{th}$ molecule. The long-term memory target data are defined on the basis of the following formula \eqref{eq2}:

\begin{equation} Y'(t)= Q^k_{in}(t-\tau) + 1/2*Q^k_{in}(t-3/2*\tau)\label{eq2}\end{equation}

$Y'(t)$ is the target data at time t. $Q^k_{in}(t)$ is the influx value of the $k^{th}$ molecule. $\tau$ indicates the number of time steps (in seconds) in the past from the current time step (past time lag).

The input signal was a random step signal generated by multiplying the fixed inflow amount with random numbers sampled from a truncated normal distribution, a normal distribution bounded between [0,1]  with a mean of 0.5 and a standard deviation of 0.25. In these input signals, the inflow values remained constant for a specified input hold time, referred to as steps. The seed value was explicitly defined for the random number generator for reproducibility. Various step sizes were tested to assess the impact of the input signal on the reservoir dynamics. The input signals were fed into both the stochastic simulation and the target data generation. The parameters for the input signals and the target data generation process are given in Section \ref{subsec:params}. For a fair training process, the stochastic data of the influx node was excluded from the training and testing process. The target data was divided into training and test sets, with 70\% used for training and the remaining 30\% for testing. These reported errors correspond to the performance of the test set.
 
\subsection{Optimizing Parameters with Genetic Algorithm} \label{subsec:params}

To automate the selection of topologies and network parameters, the genetic algorithm library called DEAP \cite{fortin2012deap} is utilized within ChemReservoir. The genetic algorithm is a heuristic method that optimizes the parameter search process comprising three key steps: selection, crossover, and mutation. In ChemReservoir, a two-level genetic algorithm approach is employed, where the outer level optimizes the topology selection and the inner level optimizes the network parameter selection process. In the first stage, the genetic algorithm searches for the best topology with the lowest normalized root mean square error (NRMSE). The objective function is the inner level of the genetic algorithm where the best parameters for the input topology are searched to give the lowest NRMSE. The system starts with a randomly generated solution population, called individuals. Starting with these individuals, each individual is used as an input of the objective function. In both levels, the population and elite sizes are set to 4 and 2, respectively. Elite selection allows storing the best 2 individuals from the former iteration of the genetic algorithm. The selection function randomly chooses three individuals from the population, and the crossover function mates two individuals. Custom mutation functions are defined at both levels. In the topology generation process, the mutation function is configured to randomly increment or decrement the number of nodes, chord length, and chord step. The mutation step size is set to 10 for the number of nodes and 2 for the chord parameters. In the network optimization process, the mutation step size is set to 0.2 for the outflow amount, 0.1 for the reaction rates, and 1 for both the inflow and the reaction scaling factors. In both levels of the genetic algorithm, the individual probability (indpb) value is set to 0.5. Individual probability is the probability that each gene (element) in an individual will be mutated. Individuals are evaluated based on the fitness score which is the NRMSE. The crossover function was set to mate two individuals.  Based on the given set of boundaries, the best topology with the best set of network parameters that returns the lowest NRMSE is searched. The boundaries of the genetic algorithms are given in Table \ref{gen-boundaries}.

\begin{table}[htbp]
  \caption{Input Parameter Boundaries for Genetic Algorithms}
  \label{gen-boundaries}
  \centering
  \setlength{\tabcolsep}{8pt} 
  \begin{tabular}{|l|l|}
    \hline
    \textbf{Parameter} & \textbf{Boundaries} \\
    \hline
    Number of nodes & (50, 300) \\
    Inflow amount & (50, 200) \\
    Chord length & (5, 25) \\
    Chord step & (5, 25) \\
    Input rate scaling & (1, 10) \\
    Reaction rate scaling & (1, 10) \\
    Outflow rates & (0.05, 1.0) \\
    Reaction rates & (0.1, 1.0) \\
    \hline
  \end{tabular}
\end{table}

In Table \ref{gen-boundaries}, the number of nodes refers to the cycle length. The inflow amount means the constant inflow value scaled by random numbers sampled from a truncated normal distribution to generate the inflow signal. The chord length is the fixed length of each chord and the chord step is the interval between nodes where chords are added. Input and reaction rate scaling values are used to evaluate the scaling factor on these rates during the genetic algorithm process. The outflow rate represents the quantity exiting the system and reaction rates are for each pseudo-rule defined as edges within the network. 

\begin{figure}[htbp]
  \centering
  \includegraphics[width=0.8\linewidth]{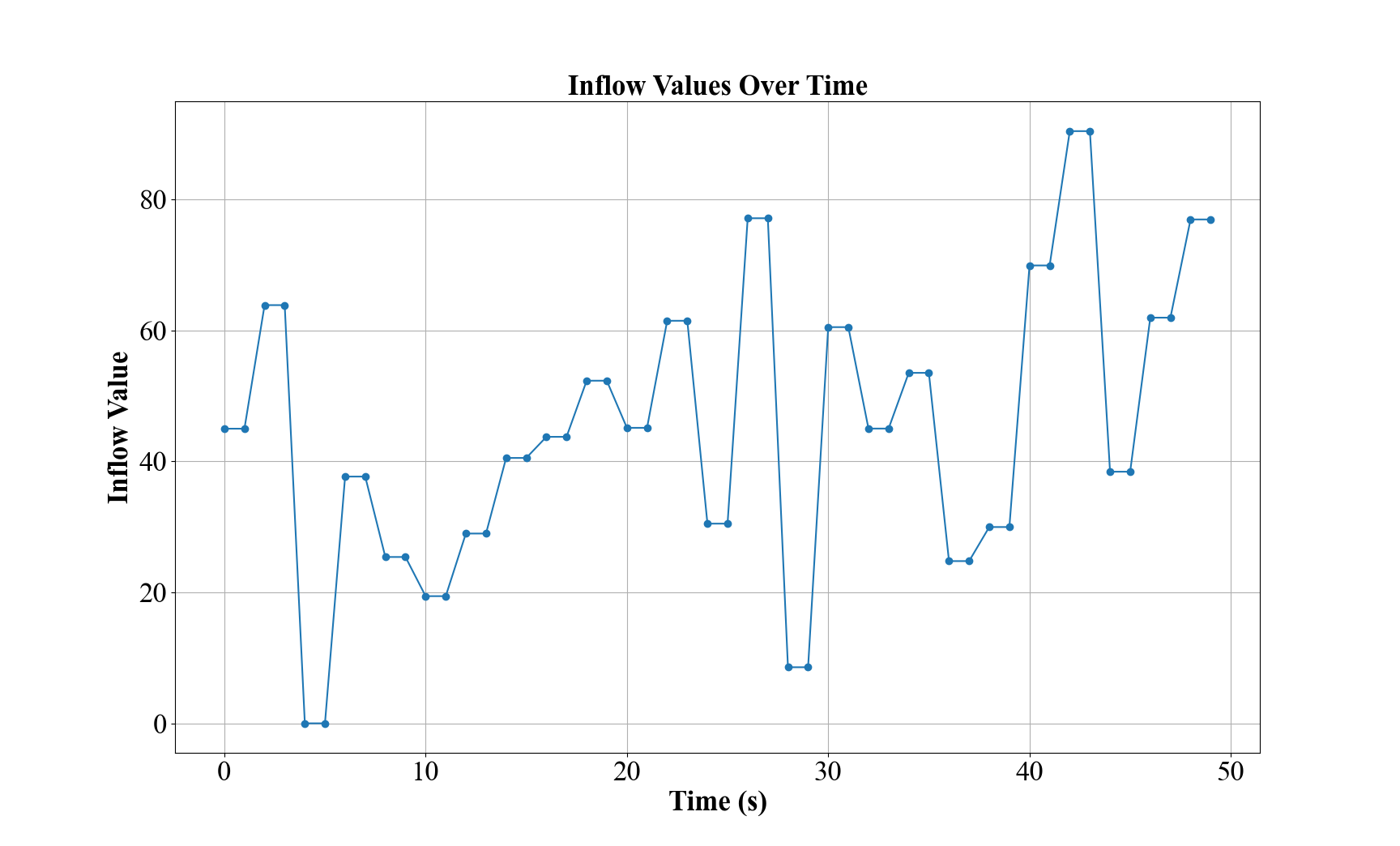}
  \caption{The inflow values over 50 s.}
  \label{fig:inflowValues}
\end{figure}

The run time for the stochastic simulation is set to 50 seconds. For these two levels of the the genetic algorithm, the maximum run times are set to 500 and 10,000 seconds, respectively. For the simulation run time, 50, the inflow random step signal is generated for various step sizes (2, 5, 10, 25). These step sizes are chosen so that the total runtime of 50 can be divided into an exact number of steps. For example, there are 25 perturbations for step size 2 in the inflow value with a total length of 50 (Fig.\ref{fig:inflowValues}). The step size performance was tested on short-term memory task and the long-term memory task was performed using the best step size. For the long-term memory task, a set of tau values, (6, 12, 18, 24), was examined. Since the inflow changes every 2 seconds, the tau values were chosen as multiples of an even number, 6, so that each delay spans multiple distinct inflow values in the past.

The ChemReservoir was run systematically for various configurations and seed value 1 was chosen as a fixed value. Due to the random number generator libraries, the seed value should be fixed for reproducibility.

\section{Results}

The benchmarking against the Nguyen's and Yahiro’s software was attempted; however, neither software is actively maintained. Therefore, only ChemReservoir results are provided. The framework was utilized for short- and long-term memory tasks. The impact of the input signal's step size was investigated for a set of step sizes (2, 5, 10, 25) for the input signal of length 50. The short-term memory task was used to investigate the impact of step size on fading memory and learning performance. For these input signals with fixed step sizes, the two-leveled genetic algorithms was utilized to optimize the topology structure and its parameters for the lowest NRMSE, as described in the previous section. Each call for the genetic algorithm started with the same initial populations for these step size values due to the fixed seed value. Table \ref{inputHoldTime-table} demonstrates the impact of step sizes on learning performance. The results indicate that the increase in the number of perturbations led to a reduction in the learning error due to temporal encoding, which is essential for the echo state property. In contrast, the infrequent changes in the input value led to fading memory as the learning performance got worse, thereby violating the conditions required for the echo state property.

\begin{table}[htbp]
  \caption{Short-Memory Task Results for Varying Step Sizes}
  \label{inputHoldTime-table}
  \centering
  \setlength{\tabcolsep}{10pt} 
  \begin{tabular}{|l|l|}
    \hline
    \textbf{Step Size} & \textbf{NRMSE} \\
    \hline
    2  & 0.348 \\
    5  & 0.532 \\
    10 & 0.936 \\
    25 & 7.342 \\
    \hline
  \end{tabular}
\end{table}

\begin{figure}[htbp]
  \centering
  \includegraphics[width=0.8\linewidth]{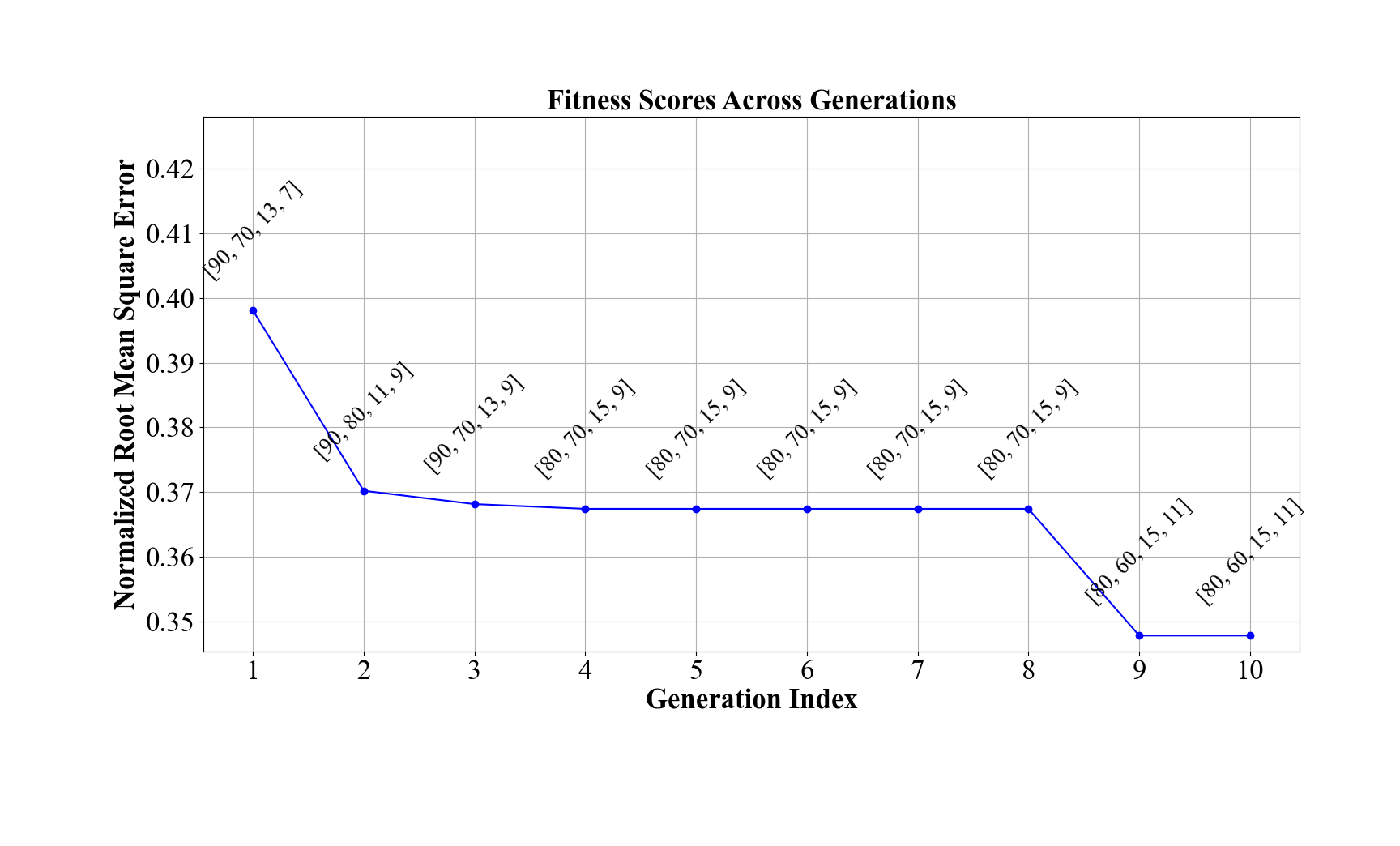} 
  \caption{Minimum fitness scores across generations for the topology optimization in the short-term memory task.}
  \label{fig:shortTopology}
\end{figure}

\begin{figure}[htbp]
  \centering
  \includegraphics[width=0.8\linewidth]{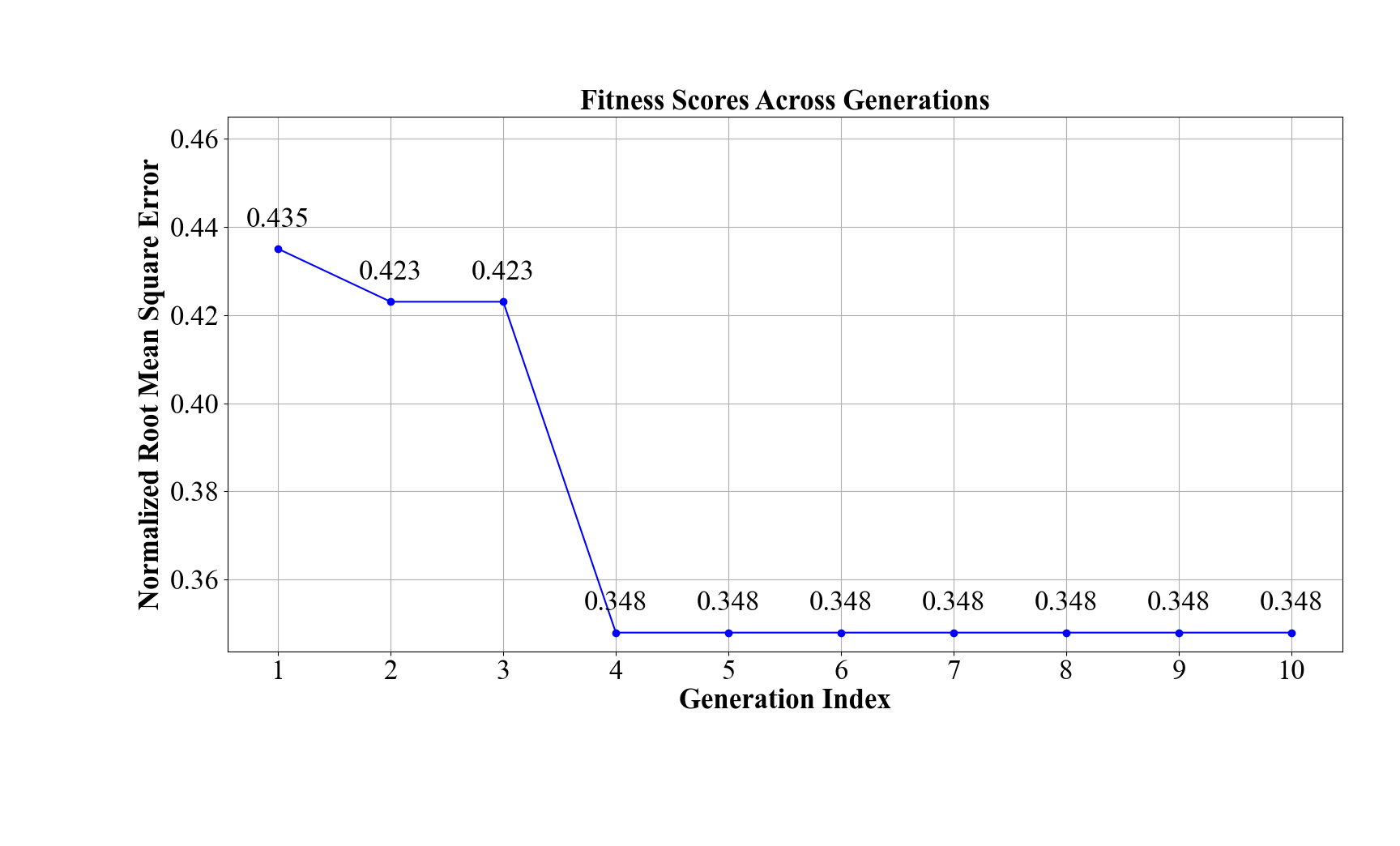} 
  \caption{Minimum fitness scores across generations for the network parameter optimization of the most optimal topology features in the short-term memory task.}
  \label{fig:shortNetworkParam}
\end{figure}

The best learning performance was obtained for the step size 2 for the short-term memory task. The results for the topology optimization and network parameter optimization are demonstrated in the Fig. \ref{fig:shortTopology} and Fig. \ref{fig:shortNetworkParam}. In Fig. \ref{fig:shortTopology}, the node labels represent the number of nodes, the fixed amount of inflow multiplied by random numbers, the chord length and the chord step, respectively. The plot shows that the decrease in chord density, the chord step value, had the main impact on learning performance. Although there were slight changes in the number of nodes, inflow amount, and chord length, the chord step was the only parameter that consistently increased. For the most optimal topology features (80, 60, 15, 11), the results of the genetic algorithm for network parameter optimization are illustrated in Fig. \ref{fig:shortNetworkParam}. The plot demonstrates that after a slight decrease during the first three generations, the NRMSE score stabilized in the fourth generation and remained the same for the remaining seven generations. 

\begin{figure}[htbp]
  \centering
  \includegraphics[width=0.8\linewidth]{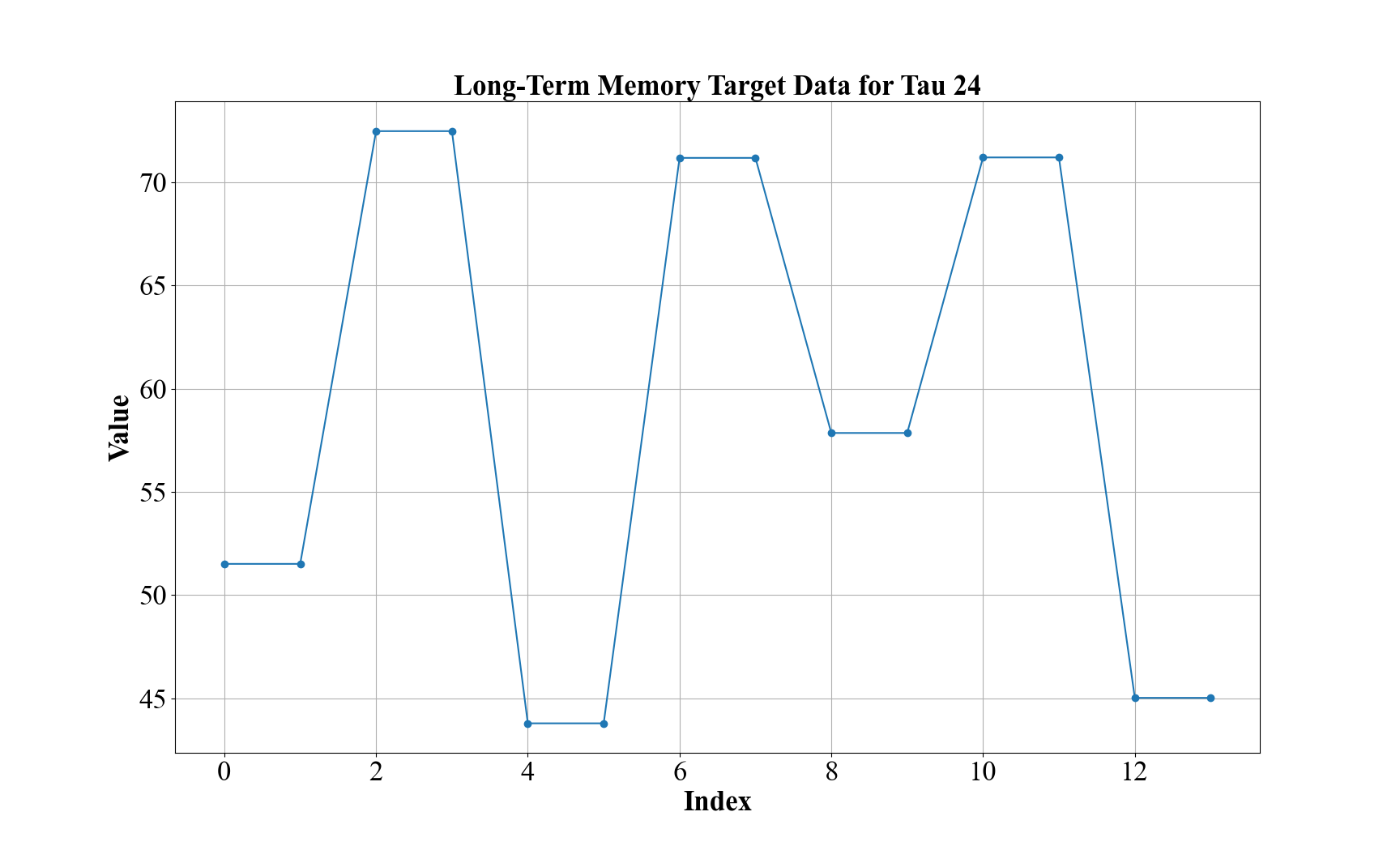}
  \caption{Long-term memory target data for tau 24.}
  \label{fig:tau24}
\end{figure}

The long-term memory task was performed for the input signal step size, 2. For this task, a list of tau values was examined. The results for these tau values are given in Table \ref{tauLong-table}. The increase in tau values, representing past time lags, led to a slight upward trend in the error values. This outcome aligns with the echo state property, which suggests that maintaining the influence of past inputs becomes more challenging as time lags increase. The only exemption was the slight decrease for tau 24; however, this was due to the symmetric structure of the target data (Fig. \ref{fig:tau24}).  

\begin{table}[htbp]
  \caption{Long-Memory Task Results for Varying Tau Values}
  \label{tauLong-table}
  \centering
  \setlength{\tabcolsep}{10pt} 
  \begin{tabular}{|l|l|}
    \hline
    \textbf{Tau Value} & \textbf{NRMSE} \\
    \hline
    6  & 0.313 \\
    12 & 0.361 \\
    18 & 0.410 \\
    24 & 0.382 \\
    \hline
  \end{tabular}
\end{table}

\begin{figure}[htbp]
  \centering
  \includegraphics[width=0.8\linewidth]{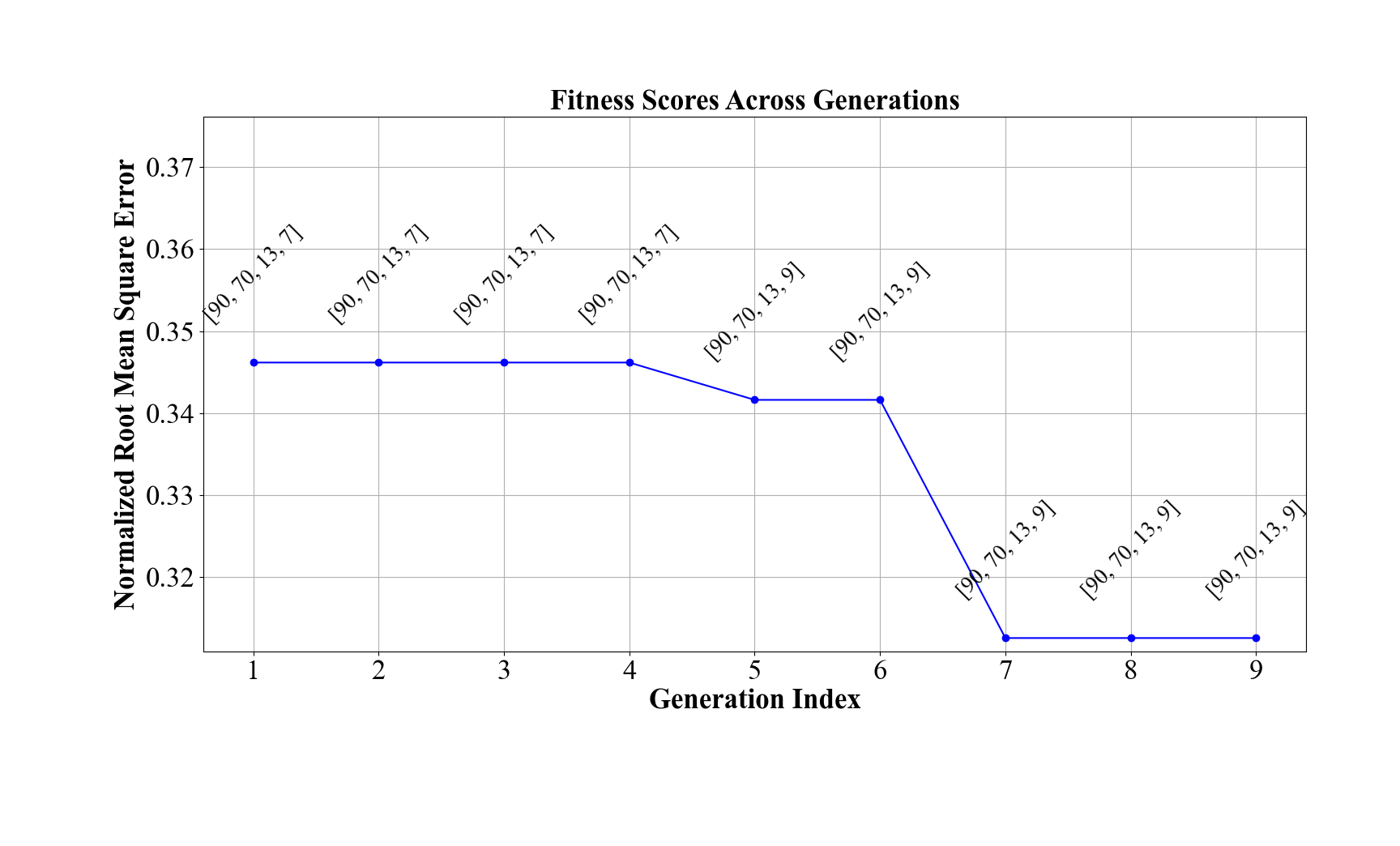} 
  \caption{Minimum fitness scores across generations for the topology optimization for tau 6 in the long-term memory task.}
  \label{fig:longTopology}
\end{figure}

\begin{figure}[htbp]
  \centering
  \includegraphics[width=0.8\linewidth]{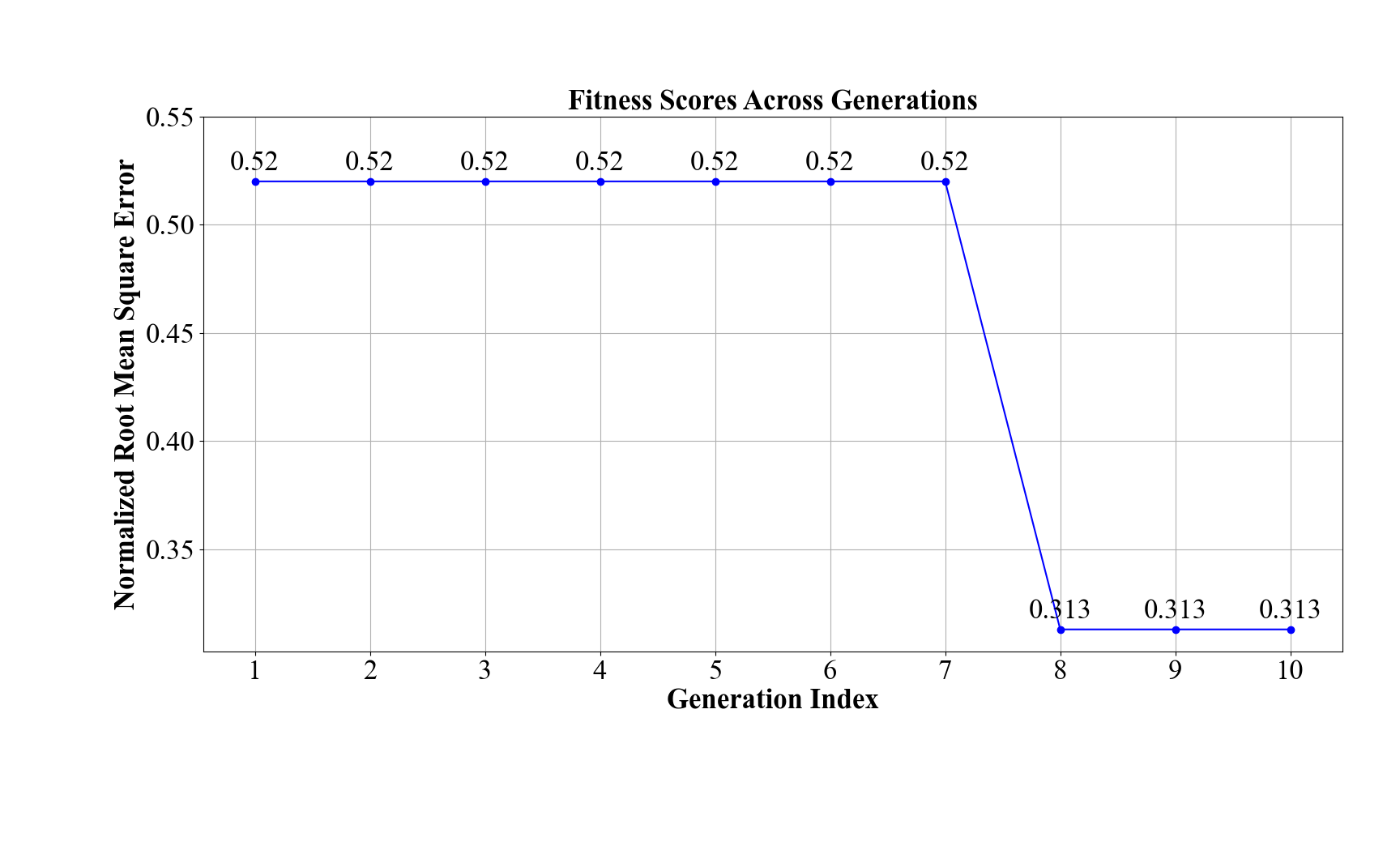} 
  \caption{Minimum fitness scores across generations for the network parameter optimization of the most optimal topology features with tau 6 in the long-term memory task.}
  \label{fig:longNetworkParam}
\end{figure}

The lowest NRMSE was obtained for the tau value 6. The changes in the minimum fitness scores for the topology optimization are illustrated in Fig. \ref{fig:longTopology}, whereas the results of the network parameter optimization for the best optimal network feature (90, 70, 13, 9) with tau 6 are shown in Fig. \ref{fig:longNetworkParam}. Compared to the short-term memory result, there were no sharp changes in the topological parameters. The lowest NRMSE was obtained for the same topology candidate for a few generations; even the decrease in NRMSE was observed for the same candidate. Therefore, topological features such as chord length and chord steps did not have a great impact compared to the short-term memory results. The increase in past time lag made the training task more complicated in handling the impact of input signal on the reservoir as expected due to fading memory.  

\section{Conclusion}

Due to the lack of actively maintained and well-tested tools for chemically-inspired reservoir computing, we developed ChemReservoir, an open-source framework designed to address this gap. Short- and long-term memory capacity tasks were performed to evaluate the memory capacity of CIR models with investigation of topological features. Similarly to the former theoretical studies on reservoir computing, our results demonstrate that the cycle-based structures on chemically-inspired reservoir models meet the echo state property, as evaluated by memory capacity tasks. From a topological point of view, the density of the local connections (chords) was examined and shown how this affects the propagation of information within reservoirs. Although modeling real chemistry is outside the scope of this work, these CIR models provide theoretical insights into the impact of input signals and local connections on the reservoir memory capacity, which may offer useful perspectives for real-chemistry implementations. 

\section{Code Availability}

The open-source framework ChemReservoir is available at:\\ 

\url{https://github.com/MehmetAzizYirik/ChemReservoir}.

\newpage

\end{document}